# Emergence of Topological Bimerons in Monolayer CrSBr

*Baishun Yang[1], Xiufeng Han[2], and Silvia Picozzi[1]*

1. Consiglio Nazionale delle Ricerche (CNR-SPIN), Unità di Ricerca presso Terzi c/o Università "G. D'Annunzio", 66100 Chieti, Italy

2. Beijing National Laboratory for Condensed Matter Physics, Institute of Physics, Chinese Academy of Sciences, Beijing 100190, China

The rich and fascinating physics of topological spin textures in van der Waals two-dimensional magnets has motivated recent growing interests, though a comprehensive understanding remains elusive. Here, in atomistic simulations on monolayer CrSBr, we find two magnetic phases emerging under non-equilibrium conditions at distinct temperatures, a standard ferromagnetic transition $T_c$ and a lower temperature $T^*$. Moreover, the real-space analysis of the spin texture reveals the emergence of metastable topological bimeron defects below $T^*$, showing an algebraic-like decaying spin-spin correlation function. The Dzyaloshinskii-Moriya interaction, induced by the local site asymmetry in the centrosymmetric CrSBr monolayer, is proved to be the origin of the bimerons formation. Furthermore, the increasing bimerons density upon increasing the cooling rate follows a Kibble-Zurek behavior, suggesting a handle to drive and control topological bimerons below $T^*$. Our results put forward CrSBr as an important candidate for the investigation of the dynamical behavior of bimerons in vdW magnets.

Van der Waals (vdW) magnets have been identified as an emerging class of materials, providing tremendous insights in fundamental condensed matter physics and potential applications in spintronics.[1-6] In particular, some nontrivial topological spin phases, such as skyrmions, merons, bubbles, etc. have been observed or predicted at equilibrium.[7-9] However, out-of-equilibrium dynamical processes, where some unconventional topological defects and domain walls might emerge,[10-12] is much less explored.

Among the recently synthesized vdW magnets, the A-type antiferromagnetic (AFM) vdW CrSBr has received great attention.[13-24] CrSBr shows high stability under ambient conditions and its magnetic transition temperature is quite high (i.e. $T_N$ = 132 K for the bulk[13] and $T_c$ = 146 K for monolayer[14]). The orthorhombic structure of CrSBr makes this material anisotropic and results in large differences in magneto-transport along the three different $a$, $b$, $c$ directions.[13, 16, 19] Even more interestingly, besides the magnetic transition temperature $T_N$, many experiments found that the magnetic susceptibility, when measured as a function of temperature, shows a kink at $T^*$ near 30 ~ 40 K.[13, 16, 18, 19] Moreover, due to the strong coupling between magnetic order and charge transport, negative magnetoresistance effects were observed at $T < T^*$.[13] The unusual changes in magnetic susceptibility and magnetoresistance suggest that an unconventional magnetic order at $T^*$ might be present, thus attracting a lot of attention. Some studies suggested that this "hidden" order comes from magnetic defects and impurities[19] or magnetic fluctuations under field[16]. Another research associated the phase transition to the slowing down of magnetic fluctuations, accompanied by a continuous reorientation of internal fields.[18] Interestingly, such an unconventional magnetic order is not observed in all experiments.[20, 21] In any case, the unresolved controversy surrounding the unconventional order in CrSBr at $T^*$ highlights the complexity of the magnetic phases of the system and urgently calls for further investigations.

In this work, we study the magnetism of monolayer CrSBr using first principles calculations, along with combined Monte-Carlo and spin dynamics simulations. Our results show that, at equilibrium, a standard FM transition occurs at 112K. However, under non-equilibrium conditions, when decreasing the temperature, the CrSBr monolayer undergoes a transition to a phase where topological bimeron-like defects form. These observations are supported by the calculations of spin-spin correlation functions and real-space spin textures. Moreover, by analyzing the phase diagrams obtained by artificially tuning the relevant parameters of the model, we found that the bimerons formation is induced by the Dzyaloshinskii-Moriya (DM) interaction, arising from the local site asymmetry in the CrSBr monolayer. Finally, the density of bimerons as a function of the temperature-quenching rate follows the Kibble-Zurek mechanism and can be further modulated by the competition between DM and exchange interactions.

The crystal structure of monolayer CrSBr is illustrated in **Figure 1a**. The lattice constants, optimized by Density Functional Theory (DFT) (see **METHODS in Supporting Information**), are $a$ = 3.506 Å, $b$ = 4.698 Å, respectively. Monolayer CrSBr belongs to the centrosymmetric *Pmmn* space group, the inversion center being located at the center of second-nearest (2[nd], label 0 and 2 in **Figure 1a**) neighbor Cr-Cr pairs. However, due to the local site inversion breaking, no inversion center occurs for the first-nearest (1[st], 0 and 1) and third-nearest (3[rd], 0 and 2) neighbor Cr-Cr pairs. Therefore, according to Moriya rules,[25] the DM interaction[25, 26] is along the ***b*** (***a***) direction and labeled as $D_y$ ($D_x$) for 1[st] (3[rd]) Cr-Cr pair in the following.

The magnetic interactions are modeled by using the spin Hamiltonian with the following general form:

$$H = \frac{1}{2}\sum_{i \neq j} S_i J_{ij} S_j + \sum_i S_i A_i S_i \tag{1}$$

where $J_{ij}$ and $A_i$ represent the exchange coupling interaction tensor and single ion anisotropy (SIA), respectively, where all the parameters can be obtained from DFT calculations using the four-state method[27, 28]. The $J_{ij}$ results are shown in **Supporting Information Table 1**. The diagonal terms $J_{xx}$, $J_{yy}$, and $J_{zz}$ of the three nearest neighbors are all positive, showing ferromagnetism and preventing magnetic frustration in monolayer CrSBr. The antisymmetric part of $J_{ij}$ corresponds to the DM interaction and can be expressed as $D_\gamma = \frac{1}{2}\varepsilon_{\alpha\beta\gamma}(J_{\alpha\beta} - J_{\beta\alpha})$, where $\varepsilon_{\alpha\beta\gamma}$ denotes the Levi-Civita symbol. The DM interactions of 1$^{st}$ and 3$^{rd}$ nearest Cr-Cr pairs in CrSBr are $D_y = 0.16$ meV and $D_x = -0.29$ meV, respectively, in agreement with the symmetry analysis by Moriya rules. Moreover, the third nearest-neighbor exchange constant, $J_{03} = (J_{xx} + J_{yy} + J_{zz})/3$, is 2.96 meV and $D_x/J_{03}$ is ~ 0.1, i.e., in the typical range for the formation of topological spin textures.[29] In addition, due to the orthorhombic structure of CrSBr, the $A_{xx}$, $A_{yy}$, and $A_{zz}$ parameters are not equal to each other; rather, the calculated $A_{xx}$-$A_{zz}$ and $A_{yy}$-$A_{zz}$ are -0.01 meV and -0.02 meV, respectively, consistent with experiments suggesting the easy, hard, and intermediate axis of CrSBr to be along the ***b***, ***c***, and ***a*** direction, respectively.[15]

Based on Eq. (1), we performed atomistic spin simulations via Monte-Carlo (MC) for a 100 nm × 100 nm monolayer CrSBr, focusing on its magnetic transition temperature. On one hand, a classic FM *M-T* curve is shown in **Figure 1b**, indicating a FM ground state of monolayer CrSBr with Curie temperature $T_c$ ~ 112K. The $T_c$ is lower than experiments, possibly due to computational parameters (such as the Hubbard U value within DFT+U) or to the underestimated lattice constant[30]. To determine the magnetic order universality class, we fit the curve using the relation $M = M_0[1 - \frac{T}{T_c}]^\beta$, where $\beta$ is the critical exponent.[31, 32] The 2D-Ising and 2D-XY models are also plotted in **Figure 1b** for comparison. The fitted $\beta$ is 0.22, suggesting that, although the monolayer CrSBr shows a (weakly) uniaxial in-plane magnetic anisotropy, its magnetic order closely follows the 2D-XY model. On the other hand, the experimentally observed phase transition at *T\** around 30 ~ 40 K is not observed in our simulations. This suggests that *T\** might be related to some metastable states, not detected by means of equilibrium MC simulations.

In order to probe the possible emergence of an unconventional order in monolayer CrSBr at *T* < *T\**, we perform non-equilibrium spin dynamics simulations via the Landau-Lifshitz-Gilbert (LLG) equation, with 10-times averaged results shown in **Figure 1c**. The in-plane magnetization, $M_{xy} = <\sqrt{(\sum_{i=1}^N S_i^x)^2 + (\sum_{i=1}^N S_i^y)^2}>/N$, is much higher than the out-of-plane component $M_z = <\sum_{i=1}^N S_i^z>/N$, indicating the easy-plane magnetism of CrSBr. However, the results of non-equilibrium spin dynamics are qualitatively different from equilibrium MC simulations, especially for the behavior at low *T* where *M/M_s* is far from 1. To have some further insights, the spin-spin correlation function, $C(r) = <S_1(r_1) \cdot S_2(r_2)>/|S|^2$, at different temperatures is evaluated by both MC (**Figure 1d**) and spin dynamics (**Figure 1e**). For MC calculations, the correlation function sharply decays from 1 at short-range to a constant at long-range for different temperatures, with increasing magnitude upon decreasing *T*. For spin dynamics simulations, the behavior of the correlation function at high $T$ (100 and 150 K) is similar to that obtained via MC. However, at intermediate $T$ (40 and 80 K), the correlation function exponentially decays to nonzero values. Interestingly, at low $T$ (0, 5, 20 K and $T < 28$ K in **Figure S1**), an algebraic-like decaying correlation function is observed, suggestive of a BKT behavior where "quasi-order" dominated by vortex-antivortex pairs may arise. We will focus on the spin dynamics results in the following, while the comparison between equilibrium MC and non-equilibrium spin dynamics calculations will be discussed at length below.

To gain a more comprehensive understanding of the magnetism at different temperatures, we further investigate the spin textures in real-space obtained with spin dynamics. The in-plane and out-of-plane magnetism are shown in the upper and lower panels of **Figure 2a**, respectively. It is clear that, upon cooling, the long-range random disordered state at 150 K (see also zoom in **Figure 2b**) changes to a nearly short-range ordered magnetic-domains state at 50 K; subsequently, some magnetic defects emerge at 10 K. Some of the defects annihilate during the cooling process and, at last, some clear magnetic defects appear to be randomly distributed at domain wall at very low temperatures (the detailed spin evolution process is shown in **Figure S2**). The representative defects in the green and orange circular regions are shown enlarged in **Figure 2c**, where in total four different topological spin textures are observed. Unlike the original 2D-XY model, where all the spins are constrained to be in-plane, our topological spin textures exhibit a 3D distribution, with spins at the core pointing out-of-plane and spins at the boundary lying in-plane. Consistent with what is expected for in-plane anisotropy, such topological defects are merons or antimerons, each having a ±1/2 topological number.[33] Interestingly, due to the conserved winding number in in-plane magnets, the core-up (core-down) antivortex and core-down (core-up) vortex always occur in pair.[34] As a result, topological defect pairs form "bimerons"[35] in monolayer CrSBr. We note here that after a long-time evolution, the bimerons disappear and the phase collapses to a trivial FM ground state. The metastable magnetic defect state is estimated to be 0.15 meV/nm$^2$ higher in energy than the FM state. The bimerons also occasionally emerge when using MC methods (see **Figure S3**).

In order to have long-range magnetic order, anisotropic magnetic interactions must be included to overcome the Mermin-Wagner theorem. To further investigate the role of the different anisotropic magnetic interactions, we artificially change the value of some interactions, with respect to the DFT-derived values considered so far. First, artificial magnetic parameters that are similar to those calculated for CrSBr (i.e., $D_x =$ -0.4 meV, $D_y/D_x = $ -0.5, $J_1 = J_2 = J_3 = J$, $J/D_x = $ -6) are used to clarify the role of SIA under an out-of-plane external magnetic field $B_z$, as shown in **Figure 3**. For magnets with easy-plane anisotropy (SIA<0), the bimerons are present under low $B_z$. It is noteworthy that these bimerons exist even at SIA=0 (see **Figure 3a, c**). When simply setting the SIA=0 and keeping other parameters as derived from DFT, the bimerons in the domain wall region are also observed (see **Figure S4**). Under a suitably applied external $B_z$, the spins will tend to rotate toward the $B_z$ direction, destroying the bimerons and forming the bubbles (cf. **Figure 3b**). When $B_z > 1$T, all the spins are along the $z$ direction and a trivial FM state occurs. Interestingly, as the conditions change from SIA < 0 to SIA > 0 (*i.e.*, easy-axis anisotropy), some domains occur at a smaller $B_z$ (see **Figure 3**). Furthermore, a suitable $B_z$ (~ 0.1 T) can change a "trivial-domains" state to a "nontrivial skyrmion bubble"[36] state (see **Figure 3d**). As expected, all spins rotate to the $z$ direction when $B_z$ is strong enough (here $B_z > 0.2$ T). Therefore, within the phase diagram shown in **Figure 3**, we can conclude that the bimerons emerge at SIA≤0 at low $B_z$. We also remark that the dipole-dipole interaction favors the in-plane spin direction in magnetic films and is reported to contribute to topological defects in $CrCl_3$.[37, 38] We note that the tiny dipole-dipole interaction in CrSBr may increase the in-plane magnetic anisotropy; however, our results show that the bimerons always occur, no matter whether the dipole-dipole interaction is explicitly considered or not (see **Figure S5**). Therefore, up to this point, neither the SIA nor the dipole-dipole interaction represent the main reason for driving the bimerons formation.

Let's now turn to another anisotropic magnetic interaction in CrSBr, *i.e.*, the antisymmetric DM interaction. Interestingly, as we artificially remove the DM interactions in CrSBr, the topological bimerons disappear, only some domains are present (see **Figure S6**). To further illustrate the role of DM interaction, we fix $D_x = -0.4$ meV, $J_1 = J_2 = J_3 = J$, SIA = 0 and focus on the variation of bimeron states (meron-antimeron pairs) as a function of $|D_y/D_x|$ and $|J/D_x|$, as shown in **Figure 4**. The bimerons emerge in all conditions in this phase diagram. Furthermore, the DM interaction will favor orthogonal spins, therefore resulting in an increase of the density of bimerons, as $|J/D|$ decreases. These results demonstrate that the antisymmetric DM interaction is crucial for the formation of topological bimerons defects in monolayer CrSBr.

As shown in **Figure S2**, the bimerons occur at the domain walls and annihilate one by one if, after cooling down the temperature, the system is allowed to evolve for a long time. Interestingly, as shown in **Figure 5a**, the density of bimerons becomes higher as the cooling rate increases. The nearly power-law relation between the density of topological defects and the cooling rate in monolayer CrSBr is suggestive of the Kibble-Zurek (KZ) mechanism.[11, 12] Though the KZ mechanism was first predicted in the early Universe and was later tested in liquid crystals, superfluid helium, superconductors, ultracold gases, etc., its investigation in magnets is still limited.[39-41] The KZ mechanism allows for the formation of defects near the phase transition; however, the cooling rate has a large influence on the evolution of the defects, suggesting that the presence of bimerons might depend on the "history" of the sample.

Finally, finite-size effects are considered. **Figure 5b** illustrates that at ~0 K, when the simulation size is smaller than 30 nm, both the in-plane and out-of-plane magnetizations show a nearly constant value and without large fluctuations, even after averaging the results 10 times. The $M_{xy}$ ~ 1 indicates monolayer CrSBr to resemble an in-plane ferromagnet for a small system size. When the system size reaches ~50 nm, the magnetization drops and, importantly, large magnetic fluctuations occur. This is attributed to the size of the system being close to the bimeron size (~38 nm, see **Section 1 in Supporting Information**), and both the bimerons and domains have the possibility to emerge when considering different initial random states (**Figure S7**). At last, when the system size is large enough, the bimerons emerge and can be uniformly distributed at the domain wall regions, leading both the averaged $M_{xy}$ and $M_z$ values to gradually decrease as the system size increases.

The equilibrium MC simulation can be approximately viewed as a non-equilibrium spin dynamics simulation with a "super-slow" cooling rate. To verify this idea, we used the output spin textures at intermediate temperature ($T_{inter}$) obtained from MC calculation as input to do the spin dynamics calculations. The total energy and evolution of the spin textures demonstrated the above idea (see **Section 2, Figure S8, and S9 in Supporting Information**). As a result, near-to-ground states at each temperature are obtained and the metastable bimeron state is hardly observed in the MC calculations (see **Figure S3**). In order to get the metastable state in MC calculations, although not following a very rigorous procedure, one can artificially decrease the equilibration steps in Monte-Carlo simulations; in that case, anomalies in the magnetic susceptibility could be observed both at $T_c$ and $T^*$ in monolayer CrSBr (see **Figure S10**), a behavior which is compatible with the experimental results. This might also explain why, besides the $T_N$, some experiments find another phase transition at $T^*$,[13, 16, 18, 19] whereas some other experiments do not[20, 21]. Moreover, we note that a different magnetic behavior is probed via neutron powder diffraction as for average magnetism and via muon spin resonance for its dynamics.[18] Although no topological bimeron defects were directly discovered, these results indicate that the equilibrium and non-equilibrium magnetic behaviors on CrSBr seem quite different.

Based on the above analysis on critical exponents, spin-spin correlation, real-space spin textures, and Kibble-Zurek-like behavior, one could speculate $T^*$ to be related to the BKT phase transition[42, 43]. We note however that our spin-Hamiltonian is markedly different from the 2D-XY model, for which the BKT physics is usually discussed. In any case, our predictions call for experimental measurements, in particular focused on detecting the above physical parameters and to shed light on the nature of $T^*$.

In summary, within the equilibrium Monte-Carlo simulations, monolayer CrSBr follows rather closely a 2D-XY model with $T_c$ = 112K. Interestingly, under non-equilibrium spin dynamics simulations, topological bimeron defects emerge at $T^*$, the density of bimerons following the KZ behavior upon temperature quenching. Finally, the spin textures of monolayer CrSBr can be tuned by modifying the SIA and/or the competition between $J$ and DM interactions. While potentially shedding light on the complex magnetization behavior observed in experiments, our results provide a picture based on topological bimeron defects possibly indicative of an unconventional BKT phase transition in low-dimensional magnets, where DM interactions play a remarkable role.

ACKNOWLEDGEMENTS
S.P acknowledges support by the Next/Generation EU program, PRIN-2022 project "SORBET: Spin-ORBit Effects in Two-dimensional magnets" (IT-MIUR Grant No. 2022ZY8HJY). X.H. acknowledges support by the National Natural Science Foundation of China [NSFC, Grant No. 12134017]. Calculations were performed at Beijing PARATERA Technology Co., LTD and at CINECA via the ISCRA C initiative (grant HP10CJ2EW2).

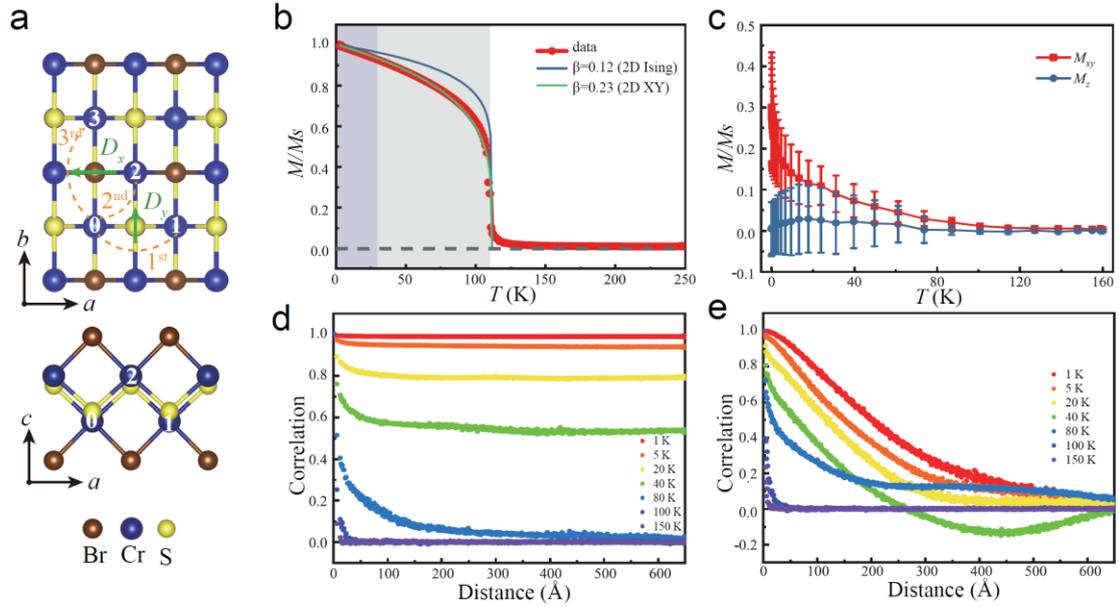

Figure 1. Schematic structure of CrSBr, atomistic simulations on M-T and correlation. (a) Top and side views of CrSBr monolayer. $M$-$T$ curve obtained by means of (b) Monte-Carlo and (c) spin dynamics simulations, the error bars for $M_{xy}$ and $M_z$ represent the dispersion obtained when averaging 10 times. The magnetic-order universality class with the 2D Ising (blue curve) and 2D-XY (green curve) models are also plotted. Spin-spin correlation function for monolayer CrSBr at different temperatures by (d) Monte-Carlo and (e) spin dynamics simulations.

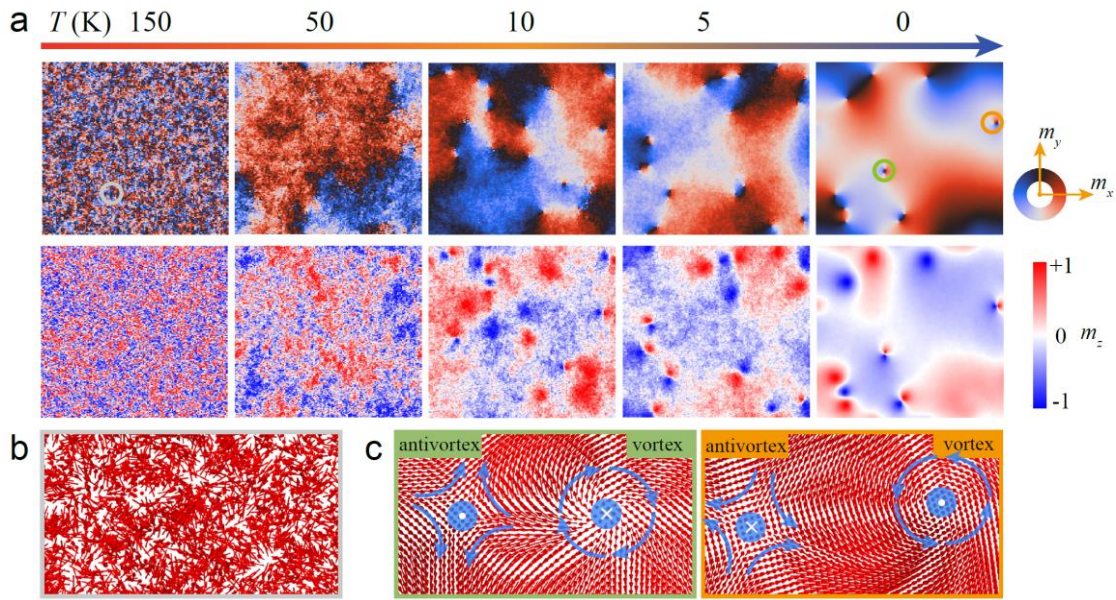

Figure 2. Real-space spin textures of CrSBr at different temperatures, obtained with spin-dynamics. (a) Selected snapshot of spin projected to in-plane (first row, see color-wheel on the right showing in-plane magnetization) and out-of-plane (second row, see color-bar on the right showing out-of-plane magnetization) as the temperature decreases from 160K to 0K. (b) Zoom-in of random spin states for the gray circle at $T$=150K. (c) Zoom-in snapshots of two different vortex-antivortex pairs for the green and orange circles at $T$~0 K. Central circles mark the out-of-plane orientation of the "core" spins.

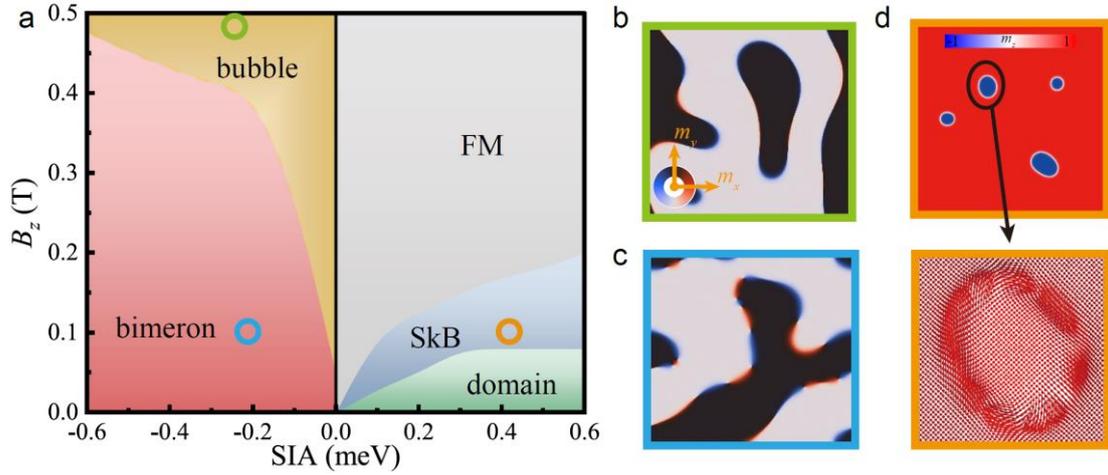

Figure 3. Phase diagram - performed with spin-dynamics - of the spin textures for varying SIA and out-of-plane magnetic field $B_z$. (a) Several spin textures - such as bimerons (red region), bubbles (yellow region), domains (green region), skyrmion bubbles (blue region), and FM (gray region) - are present in the phase diagram. Enlarged (b) bubble states, cf. green circle, (c) bimeron states, cf. blue circle, and (d) skyrmion bubble states, cf. yellow circle, respectively.

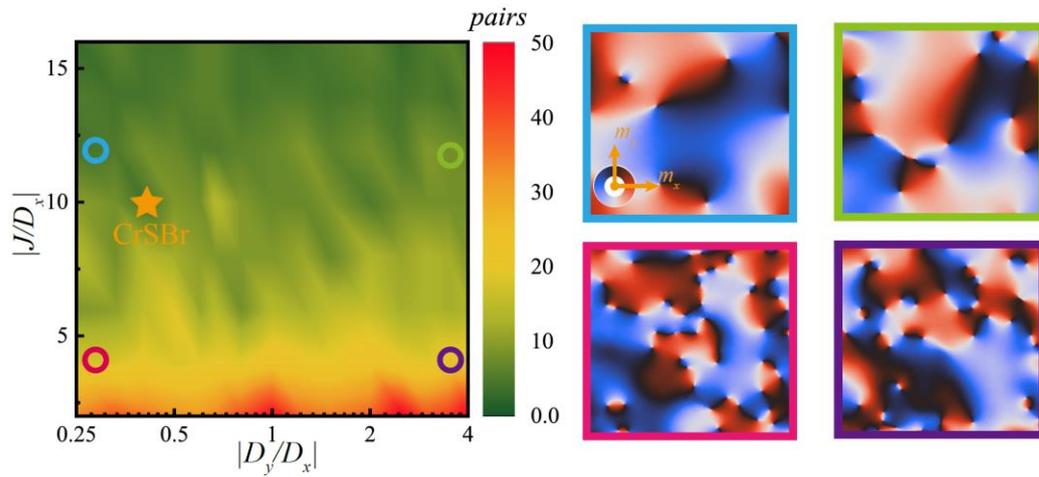

Figure 4. Phase diagram of meron-antimeron pairs as a function of $|D_y/D_x|$ and $|J/D_x|$ obtained by spin-dynamics. The enlarged in-plane spin projection for the blue, red, green, and purple circles are shown on the insets (right panels).

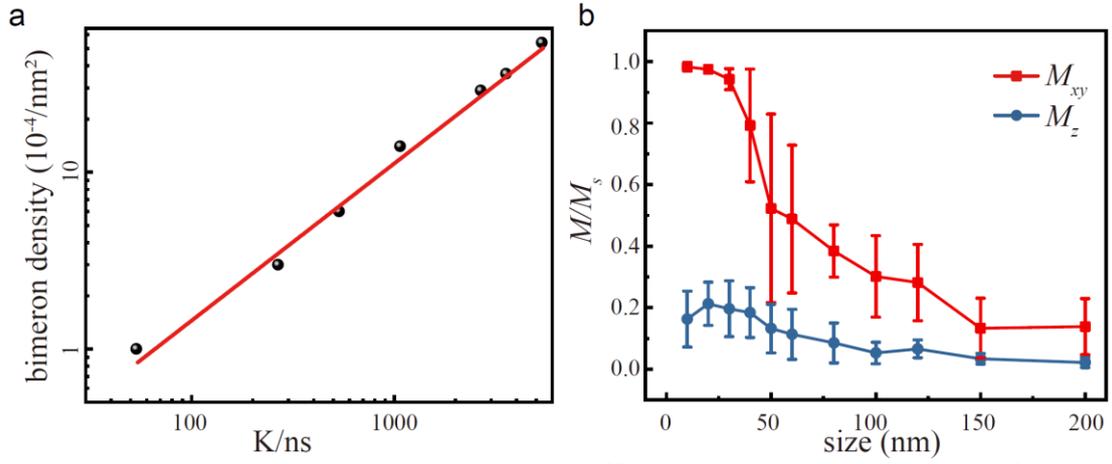

Figure 5. Kibble-Zurek mechanism and size effect. (a) Bimeron density as a function of cooling rate, following a linear relation in logarithmic coordinates. (b) The averaged in-plane (red curve) and out-of-plane (blue curve) magnetic moment for CrSBr for different system sizes. The error bars for $M_{xy}$ and $M_z$ represent the dispersion obtained when averaging 10 times.